\documentclass[sigconf,authorversion]{acmart}
\AtBeginDocument{%
  }

\setcopyright{acmlicensed}
\copyrightyear{2025}
\acmYear{2025}
\acmDOI{XXXXXXX.XXXXXXX}
\acmConference[SC25]{}{November 16-21, 2025}{St. Louis, Missouri, USA}
\acmISBN{978-1-4503-XXXX-X/2025/06}

\begin{document}

\title{AI Data Centers Need Pioneers \\ to Deliver Scalable Power via Offgrid AI}

\author{Steven P. Reinhardt}
\email{steve@xfr.ai}
\orcid{0000-0003-4355-6693}
\affiliation{%
  \institution{XFR (Transform Computing, Inc.)}
  \city{Eagan}
  \state{MN}
  \country{USA}
}








\renewcommand{\shortauthors}{Reinhardt}

\begin{abstract}
The scalable computing revolution of the late '80s through mid-'00s forged a new technical and economic model for computing that delivered massive societal impact, but its economic benefit has driven scalability to sizes that are now exhausting the energy grid's capacity.
Our time demands a new revolution in \emph{scalable energy}, mirroring in key ways the scalable computing revolution; e.g., compelling economic forces,  use of mass-market components,
overcoming foibles of those components,
judicious use of physical locality, and the 
the difficult integration into an effective system.
The \emph{offgrid AI} approach closely fits this mold, combining local mostly renewable generation and storage to power an AI data center, starting offgrid.
Obstacles to delivering this approach are social, technical, and project, but the potential is massive.
I argue that the offgrid-AI approach needs pioneers among both system developers and AI-data-center operators to move it quickly from concept to large-scale deployment.
\end{abstract}




\received{12 August 2025}
\received[revised]{12 March 2049}
\received[accepted]{5 June 2049}

\maketitle

\section{Introduction}
In 1985, supercomputing was dominated by monolithic systems designed from the ground up (e.g., processor architecture, I/O, and compilers) for that market, systems that were delivering increased computing power roughly as predicted by Moore's Law.
Despite that success, computing pioneers saw the need for dramatically greater amounts of compute power applied to a wider range of problems and  envisioned and began advocating for \emph{scalable computing}.  
Key tenets of scalable computing included the requirement to scale system size to match the problem; the use of mass-market components; and acceptance of the need to rewrite most system and application  software to accommodate the accompanying shift from shared memory to distributed memory.

By 1996, the Cray T3E system was delivered and widely recognized as the first production scalable computer.  In 2000, the last shared-memory vector Cray T90 product  shipped its final machine.  By 2005 the last vector Cray X1e machine  shipped and the switch to mass-market scalable computing was complete.

Since then, the promise of scalable computing has been fulfilled perhaps beyond its pioneers' wildest dreams, as \emph{hyperscale} systems at the biggest tech companies are much larger than the biggest science-and-engineering systems, and in the last few years massive scalable systems targeted at AI training also dwarf the biggest science-and-engineering systems (e.g., xAI's Colossus AI-training system is reported to have 200,000 GPUs while Lawrence Livermore National Lab's El Capitan system, \#1 on the June 2025 Top500 list of science-and-engineering systems \cite{Top500_2025jun}, has ~44,000 GPUs).

\begin{figure}[ht]
  \centering
  \includegraphics[width=\linewidth]{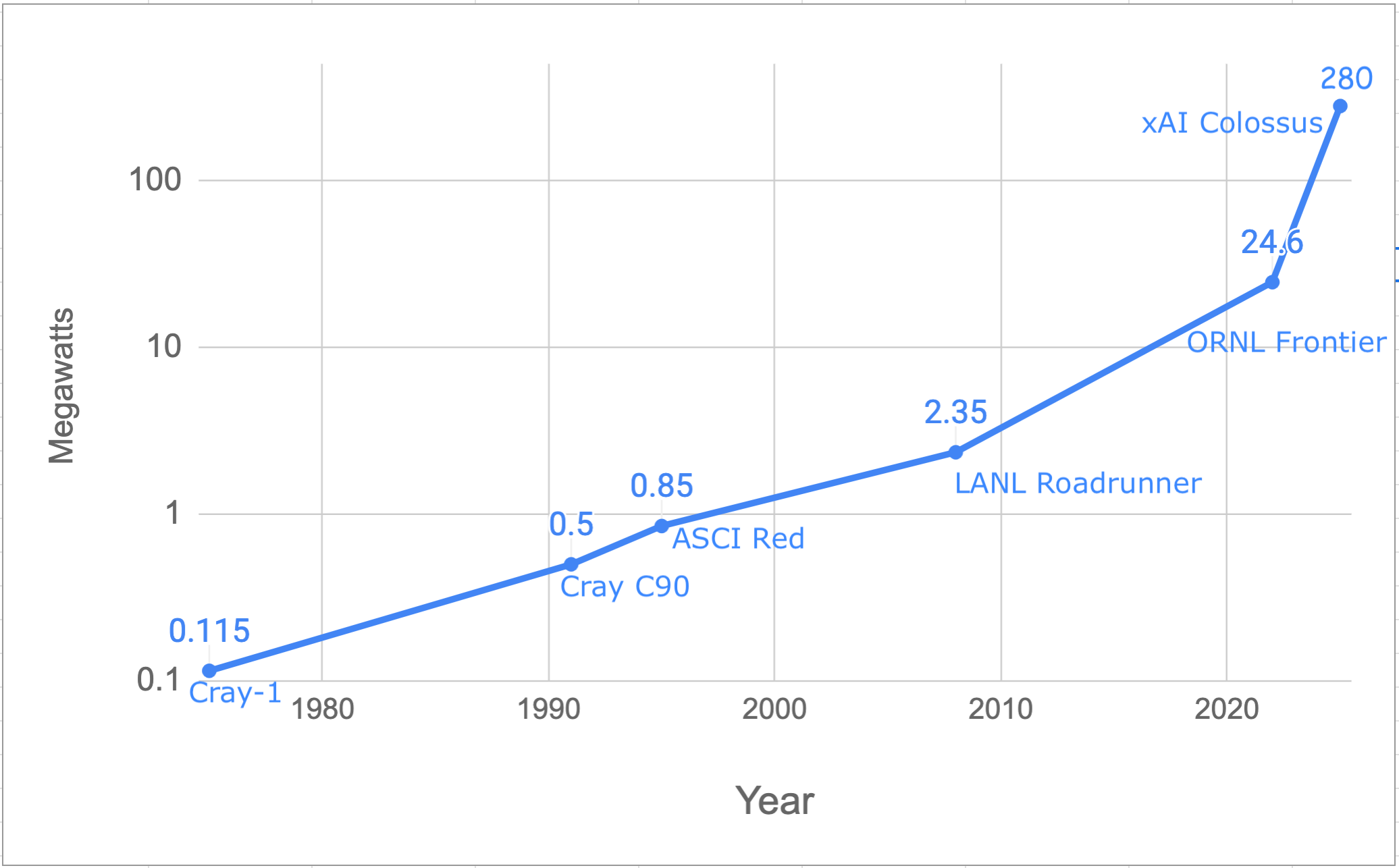}
  \caption{Power consumption of supercomputers, 1975-2025.}
  \label{fig:hec_power_consumption}
  \Description{Chart showing the growth in electricity consumed by high-end computers from 1975 to 2025}
\end{figure}

Despite intense design focus for the last 20 years on more energy-efficient computing, this relentless scaling of computing has resulted in the power consumption of the largest systems growing by over a factor of 1000, from 115kW for a Cray-1 (1975) 
to a reported 280MW for Colossus (2025) \cite{Pilz_Trends_AI_Supercomputers}, as shown in Figure \ref{fig:hec_power_consumption}.
(Computational power (in 64-bit FLOPS) between those two systems has grown by a factor of  \textasciitilde10\textsuperscript{19}, so there have been drastic improvements in ops/watt.) 
While one such system could conceivably be served by the existing energy grid, the total new power required by hundreds of such AI data centers is exhausting the grid's generation and transmission capacities and forcing both grid operators and AI-data-center operators to think more creatively about how to supply the estimated 100 GW of new power needed by 2030, a ~20\% increase in US electricity generation.

I helped drive the industrial shift to scalable computing via leading the development of Cray T3E and SGI Altix systems, and so the parallels between this current need for \emph{scalable energy} and the 1980s' need for scalable computing are striking to me:  economic forces are demanding the scalability be delivered; mass-market components that supply the core resources (microprocessors and DRAM then, solar panels and batteries now) are available, with market forces expected to dramatically improve their effectiveness and cost-effectiveness;  those mass-market components were not designed with centralized scalable systems in mind and thus  lack capabilities that would make integration easier; judicious use of physical locality can side-step or minimize issues that are expensive or impossible to overcome directly (scalable shared memory then; unavailable spare transmission grid capacity now); and lastly, a conceptually simple integration task will not be simple in practice, requiring software to address problems  not yet encountered.

Baranko et al. \cite{Baranko_Fast_Scalable_Clean_Cheap} first publicly articulated and popularized the \emph{offgrid AI} model. 
Now multiple groups working on generating power for AI data centers advocate offgrid AI, which responds to the lack of spare US transmission grid capacity by shifting offgrid the entire complex of energy generation, storage, and load  at least initially.  
Offgrid AI colocates mostly (say 90\%) solar generation with modest  (say 10\%) natural-gas generation (to keep battery costs reasonable) with batteries, control electronics,  coordinating software, and the AI data center to deliver new power quickly, reliably, resiliently, cheaply, and cleanly.
While a compelling idea, offgrid AI is today an unproven approach that needs validation in the real world.  
In this paper, I advocate for that proof and the steps to get it.

The rest of the paper briefly captures the current situation in Section 2, describes offgrid AI in Section 3, sketches some of the key obstacles to offgrid AI's maturation in Section 4, describes the needed steps to maturity in Section 5, and ends with a call to action in Section 6.

\section{Acquiring New Power Today}
The energy grid before the widespread adoption of renewable energy consisted of hundreds of generating plants serving hundreds of millions of loads, with power flowing one direction from generation to load (consumption).
Utilities and grid operators controlled the generation and distribution of power.
The amount of electricity consumed grew slowly or even not at all for many years, and the energy industry responded to society's requirements of reliability and low cost with minimal change or spare capacity (either generation or transmission).

The advent of widespread solar and wind generation, esp. in small configurations like residential rooftop solar, has changed the number of generation points to roughly equal the number of load points on the grid, at the same time that power now flows both to and from locations that previously only consumed power.
Solar and wind are both variable, unlike conventional generating plants, meaning that their generation must be more actively mediated to meet loads. 
Energy storage, usually batteries, is expensive and short-duration at utility scale.
The lack of spare generation capacity means that many utilities cannot quickly fulfill requests for new power.
Another important complication is that the lack of spare transmission capacity means that a new load must not only find new generation but must find new generation in the right location. 

The recent explosion in data centers, esp. for AI, erupted into this already dynamic situation, exacerbating an already fraught energy grid (witness power outages in California due to wildfires and in Texas due to Winter Storm Uri).  
The risks and opportunities have all been magnified.

\section{\emph{Offgrid AI} As a Guidepost}
\begin{center}
Sherlock Holmes: \emph{"When you have eliminated the impossible, whatever remains, however improbable, must be the truth."}
\end{center}

\begin{figure}[ht]
  \centering
  \includegraphics[width=\linewidth]{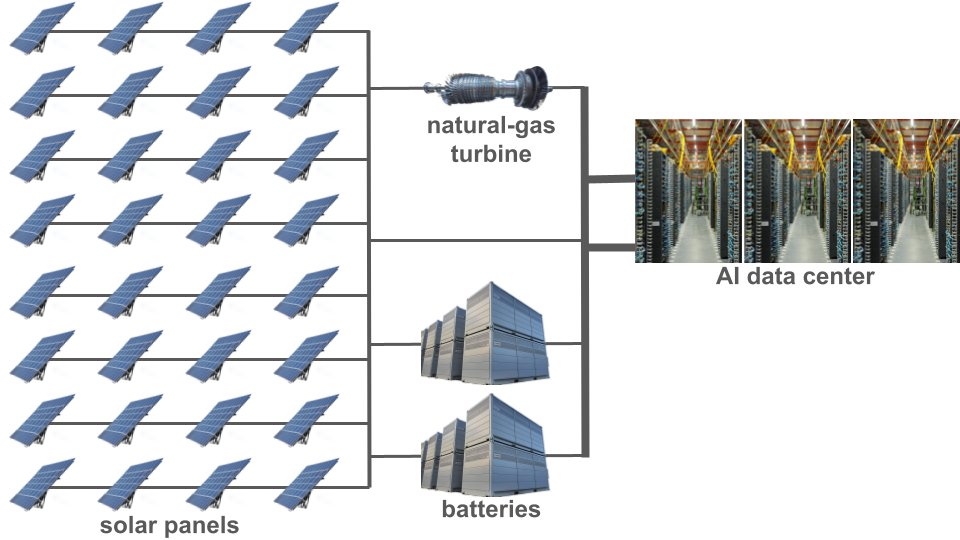}
  \caption{The offgrid-AI architecture.}
  \label{fig:offgrid_AI_architecture}
  \Description{Solar panels, natural-gas turbine, batteries, and an AI data center connected in an offgrid-AI configuration.}
\end{figure}

To address the current demand for AI-data-center power quickly and reliably, the offgrid-AI approach prescribes ~90\% solar generation and ~10\% natural-gas generation; the natural-gas component limits the long tail of the solar-generation probability curve and hence saves substantially on the lithium-ion batteries that provide storage.  
Energy generation, storage, and load (the AI data center) are physically colocated and electrically connected to each other, but without a connection to the greater grid.
Figure \ref{fig:offgrid_AI_architecture} illustrates a diagram of the electrical connectivity of an offgrid-AI site.

Note that offgrid-AI enables many other configurations if/when they become technically and economically feasible.  
Renewable generation could come from wind mills or a combination of wind mills and solar panels, exploiting their distinct variabilities, instead of only solar panels.  
(Renewable generation could also come from geothermal or hydro, but those nearly constant or \emph{firm} energy sources may not need the offgrid-AI machinery for power smoothing the way solar and wind do.)
Different energy storage technologies could be substituted, and even extending the architecture to include short- and long-duration storage could be high value.
The offgrid aspect of this would likely change over time, as there are many benefits to being grid-connected, as detailed by Gimon et al. \cite{Gimon_Energy_Parks}, despite the current big negative of long delays due to interconnection queues.  
Near-term, deferring some technical and regulatory issues by being offgrid enables getting the core technology working well faster.

\begin{figure}[ht]
  \centering
  \includegraphics[width=\linewidth]{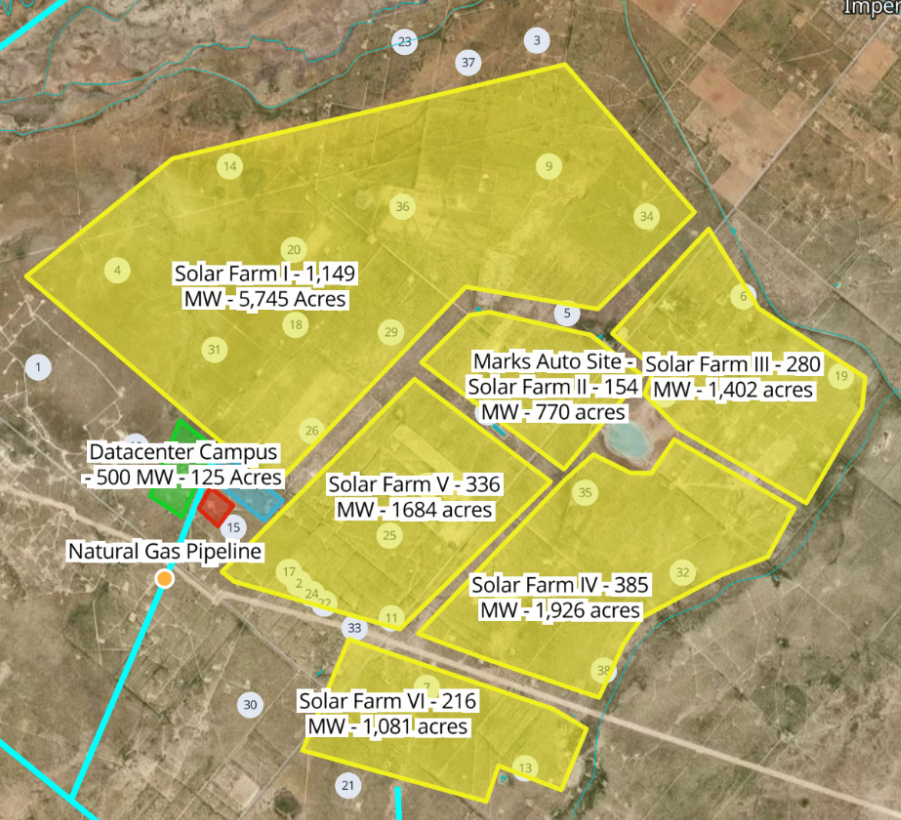}
  \caption{Diagram showing the offgrid-AI architecture.}
  \label{fig:Fast_Scalable_Clean_image6}
  \Description{Solar panels, natural-gas turbine, batteries, and an AI data center connected in an offgrid-AI configuration.}
\end{figure}

A simple diagram like Figure \ref{fig:offgrid_AI_architecture} does not convey the physical magnitude of a large offgrid-AI installation.  Figure \ref{fig:Fast_Scalable_Clean_image6} (cropped figure from \cite{Baranko_Fast_Scalable_Clean_Cheap}) illustrates an offgrid-AI installation consuming 500MW of power continually (i.e.,  \textasciitilde360,000 NVIDIA H100 equivalents).  
The yellow solar farm areas total 19.7 square miles, i.e., the figure is \textasciitilde5 miles on each side.  
By contrast the area for natural-gas generation is 0.04 square miles and for batteries is 0.12 square miles.
For comparison, the land area of the island of Manhattan is 22.8 square miles.
Despite the significant land consumed by this example, Baranko et al. \cite{Baranko_Fast_Scalable_Clean_Cheap} found that the southwest US contains  land for more than 1,200 GW of new power suitable for offgrid-AI use.

\subsection{Advantages}
\begin{itemize}
    \item Power and data centers in the offgrid-AI model are deliverable faster than if using other power sources (utility power, grid connection) and at massive scale.
    \item The raw materials are widely available and proven.  Solar irradiance is widely available, and solar panels and batteries are widely available to large markets that drive manufacturing economies of scale.
    \item Solar panels and batteries have a levelized cost of energy (LCOE) that is only slightly above natural-gas turbines today and is expected to be cheaper than natural gas by 2030 \cite{dnv_energy_transition}.
    \item Given the current impacts of the climate crisis (e.g., extreme flooding,  wildfires causing cross-continental air-quality warnings) and the factual fragility of climate denial, popular opinion could surge strongly in favor of renewable resources, rendering unclean technologies undeployable in practice, a big issue for energy equipment expected to last 30 years.
\end{itemize}

\subsection{Disadvantages}
\label{subsec:Disadvantages}
\begin{itemize}
    \item Unproven integrating technology.  The offgrid approach is conceptually appealing, but has not been implemented and proven at scale for a demanding, high-profile industry that must plan several years into the future.  Specifically, operating software and analytics enabling reliable and resilient operation do not yet exist, and existing software questionably scales to the expected installation sizes.
    \item Data-center operators are accustomed to their power coming from utilities and being firm and predictable.  Basing their data center's operation on an inherently variable power source is strongly counterintuitive bordering on anathema to many of them.  Put another way, they will require ironclad evidence that the offgrid-AI approach works in practice before committing to use it.
    \item Lithium-ion batteries are expensive and short duration to cover the \textasciitilde18 hours without sun in many locations.
    \item The primary components -- solar panels, batteries, small natural-gas turbines, and other electrical equipment such as inverters -- may have their own supply-chain issues even if not as severe as large natural-gas turbines.
    \item Solar irradiance varies considerably between southern and northern latitudes, and offgrid AI's effectiveness will be significantly reduced at northern latitudes.
\end{itemize}

\section{Technical and Project Obstacles}

\subsection{Data-center operators' discomfort with variable power sources}
As noted above in Section \ref{subsec:Disadvantages}, data-center operators have always gotten firm power from utilities and are uncomfortable with basing their data center's operation on an inherently variable power source.  
It may be that they have no other choice, but this discomfort may hamper the fine-grained cross-organization collaboration between equipment suppliers and data-center operators required to make offgrid AI a production capability.

\subsection{Interacting with regulators}
Unlike the scalable computing shift, where the physical changes occurred almost completely within the bounds of the data center proper, the shift to scalable energy will require major coordination with external bodies, notably energy regulators.
The offgrid-AI approach reduces those interactions initially by being offgrid and hence subject to less regulation, but some regulation will still be in place in many jurisdictions.  
Energy regulators are not used to changing at the often-breakneck pace of the computer industry and, given the huge responsibility to deliver reliable power, are not typically adherents of the "move fast and break things" approach sometimes used in the computer industry.

\subsection{Device coordination at scale}
The scale of expected installations is staggering; e.g., the solar farms represented in Figure \ref{fig:Fast_Scalable_Clean_image6} would require roughly 5 million solar panels.
Solar farms at this scale have encountered major issues causing curtailment by grid operators, notably electromagnetic transients (in California, Texas, and Australia)  \cite{NERC_Odessa_Disturbance} and protection coordination \cite{saeed2024utilities}.  
The single-purpose nature of offgrid AI may avoid having to solve these issues in their full generality, but they will have to be reliably solved to deliver high-quality power to the AI data center.

\subsection{Orchestration}
In addition to scale, as the collection of devices grows more varied (e.g., solar panels from multiple manufacturers with different failure characteristics, support for both short- and long-duration storage) and the reliability of offgrid-AI sites rises, the remaining issues to improve (e.g., optimizing over-/under-charge for batteries, minimizing natural-gas use) will become more difficult and complex.
At XFR, we see that analytics for offgrid-AI installations need to be markedly better than conventional approaches, as installations will be much bigger and more dynamic than previously and be required to deliver high reliability and resilience  quickly.
Addressing known issues effectively and providing general situation awareness, both electrical and cybersecurity, to isolate emergent issues quickly will be essential.

\subsection{Cost of testbeds for validation}
By my estimate, the cost to construct and populate an offgrid-AI configuration consuming 500MW of firm power would be \$25-30B.  
Even the largest equipment manufacturers will not be able to afford such a configuration for testing, so intensive small-scale and virtual testing combined with collaborative large-scale testing on customer equipment will have to play a major role. 

\subsection{Cybersecurity defense}
Super-sized AI data centers costing tens of billions of dollars to construct and populate and generating customer value several times that will be among the most inviting cybersecurity targets in existence, both to criminals and to antagonistic nation states.
Protecting offgrid-AI sites from cyber attacks, esp. given that low-level device communications (e.g., SCADA) are inherently insecure, will be a necessary and difficult task.
Note that this risk exists for grid-connected solar and battery farms just as it does for offgrid sites.

\section{Early and Needed Proof Points}
Robust implementations of offgrid AI do not exist today, and considerable work will be required to implement it at the levels of reliability and resilience required by data-center operators.  This section sketches where the community is today and steps that might convince data-center operators of offgrid-AI's maturity.

To my knowledge, there are no data centers today operating with the offgrid-AI approach, though my information may be incomplete.  
There are data centers described as running solely from renewable energy, such as Apple's \cite{Skidmore_Apple_renewable_data_center_power}, which use renewable energy generated off-site, are grid-connected,  and do not address the difficulties of supplying firm power purely locally, and Moro Hub's Green Data Center \cite{Nair_Moro_Hub_green_data_center}, which was announced as a solar data center, including batteries, but few details and little operational information is available.

Energy grids, even small offgrid ones, are dynamic systems.  
Beyond ensuring a system runs reliably in ostensibly steady state, the value in testing and validating is encountering unusual or end-case situations where the grid may not respond effectively.  
The probability of seeing such events can be increased by modifying the load, the duration of the test, the size of the grid,  the configuration of the grid, and other aspects of the grid and test.
    
Possible milestones could include those below, starting with non-positively numbered ones that use less than the full offgrid-AI configuration:

\begin{enumerate}
    \setcounter{enumi}{-2}
    \item Operate a utility-scale solar farm, specifically focusing on known issues that have disrupted solar-farm operation and so will carry forward into the solar farm  part of an offgrid-AI configuration.
    \item Operate an offgrid-AI configuration with the nominal AI-data-center load replaced by  (e.g.) crypto-coin miners.
    \item\label{item:small_ogAI} Operate a small (say 1MW) offgrid-AI configuration reliably for several months.
    \item\label{item:small_dynamic_ogAI} Hardening Item \ref{item:small_ogAI}, operate the same configuration but with anomalies (e.g., wild swings in power load, injected component failures or inactivations, insufficient generation or battery provided for the given load) provoked.
    \item\label{item:medium_ogAI}  Operate a larger (say 10MW) offgrid-AI configuration reliably for several months.
    \item\label{item:medium_dynamic_hardened_ogAI}  Operate the same offgrid-AI configuration of Item \ref{item:medium_ogAI}, with the complications of Item \ref{item:small_dynamic_ogAI}.
    \item\label{item:large_ogAI}  Operate a larger (say 100MW) offgrid-AI configuration reliably for several months.
    \item\label{item:large_dynamic_hardened_ogAI}  Operate the same offgrid-AI configuration of Item \ref{item:large_ogAI}, with the complications of Item \ref{item:small_dynamic_ogAI}.
    \end{enumerate}


\section{Call to Action}
Delivering scalable computing's promise for future decades requires scalable energy.
The offgrid-AI approach to scalable energy offers a clear path, with risk, to deliver the massive amounts of new power needed quickly, reliably, resiliently, cheaply, and cleanly.
As Baranko et al. say so well, "If this is so great, why isn’t it happening? ... But the biggest reason may simply be inertia and the fact that this hasn’t been done before."
For this transformation to happen as quickly as needed, the following steps need to happen.
\begin{itemize}
    \item Offgrid-AI advocates must implement near-term proof points  aggressively to build confidence in AI-data-center developers.
    \item Pioneers in the AI/scalable-computing  and  energy worlds must grok the offgrid-AI potential and connect with each other across disciplines.
    \item Pioneers in the AI-data-center world must foster offgrid-AI technology at smaller scales and lower interim reliability/resilience levels.
    \item Funders, both industrial and governmental, must see the chasm that must be crossed and infuse funds for proof points.
    Smaller investments than those already made by hyperscalers in small modular nuclear and geothermal power will prove the offgrid-AI approach.
\end{itemize}
With these steps, scalable computing and scalable energy can deliver economic value  to the next generation.

\bibliographystyle{ACM-Reference-Format}
\bibliography{scalable_energy}


\begin{thebibliography}{9}


\ifx \showCODEN    \undefined \def \showCODEN     #1{\unskip}     \fi
\ifx \showISBNx    \undefined \def \showISBNx     #1{\unskip}     \fi
\ifx \showISBNxiii \undefined \def \showISBNxiii  #1{\unskip}     \fi
\ifx \showISSN     \undefined \def \showISSN      #1{\unskip}     \fi
\ifx \showLCCN     \undefined \def \showLCCN      #1{\unskip}     \fi
\ifx \shownote     \undefined \def \shownote      #1{#1}          \fi
\ifx \showarticletitle \undefined \def \showarticletitle #1{#1}   \fi
\ifx \showURL      \undefined \def \showURL       {\relax}        \fi
\providecommand\bibfield[2]{#2}
\providecommand\bibinfo[2]{#2}
\providecommand\natexlab[1]{#1}
\providecommand\showeprint[2][]{arXiv:#2}

\bibitem[Baranko et~al\mbox{.}(2024)]%
        {Baranko_Fast_Scalable_Clean_Cheap}
\bibfield{author}{\bibinfo{person}{Kyle Baranko}, \bibinfo{person}{Duncan Campbell}, \bibinfo{person}{Zeke Hausfather}, \bibinfo{person}{James McWalter}, {and} \bibinfo{person}{Nan Ransohoff}.} \bibinfo{year}{2024}\natexlab{}.
\newblock \bibinfo{title}{Fast, scalable, clean, and cheap enough: How off-grid solar microgrids can power the AI race}.
\newblock \bibinfo{howpublished}{\url{https://www.offgridai.us/}}.
\newblock
\newblock
\shownote{Accessed: 2025-08-10}.


\bibitem[{DNV}(2023)]%
        {dnv_energy_transition}
\bibfield{author}{\bibinfo{person}{{DNV}}.} \bibinfo{year}{2023}\natexlab{}.
\newblock \bibinfo{title}{{DNV Energy Transition Outlook 2023}}.
\newblock
\urldef\tempurl%
\url{https://www.dnv.com/energy-transition-outlook/download}
\showURL{%
\tempurl}
\newblock
\shownote{Last accessed 2024 August 12}.


\bibitem[Gimon et~al\mbox{.}(2024)]%
        {Gimon_Energy_Parks}
\bibfield{author}{\bibinfo{person}{Eric Gimon}, \bibinfo{person}{Mark Ahlstrom}, {and} \bibinfo{person}{Mike O’Boyle}.} \bibinfo{year}{2024}\natexlab{}.
\newblock \bibinfo{title}{Energy Parks}.
\newblock \bibinfo{howpublished}{\url{https://energyinnovation.org/wp-content/uploads/Energy-Parks-Report.pdf}}.
\newblock
\newblock
\shownote{Accessed: 2025-08-10}.


\bibitem[Nair(2024)]%
        {Nair_Moro_Hub_green_data_center}
\bibfield{author}{\bibinfo{person}{Arya~M Nair}.} \bibinfo{year}{2024}\natexlab{}.
\newblock \bibinfo{title}{Saeed Al Tayer reviews 2nd phase expansion of Moro Hub green data center}.
\newblock \bibinfo{howpublished}{\url{https://www.gccbusinessnews.com/saeed-al-tayer-reviews-green-data-centers/}}.
\newblock
\newblock
\shownote{Accessed: 2025-08-12}.


\bibitem[Pilz et~al\mbox{.}(2025)]%
        {Pilz_Trends_AI_Supercomputers}
\bibfield{author}{\bibinfo{person}{Konstantin~F. Pilz}, \bibinfo{person}{James Sanders}, \bibinfo{person}{Robi Rahman}, {and} \bibinfo{person}{Lennard Heim}.} \bibinfo{year}{2025}\natexlab{}.
\newblock \showarticletitle{Trends in AI Supercomputers}.
\newblock \bibinfo{journal}{\emph{arXiv preprint}} \bibinfo{volume}{2025}, \bibinfo{number}{16026} (\bibinfo{date}{April} \bibinfo{year}{2025}).
\newblock
\urldef\tempurl%
\url{https://arxiv.org/pdf/2504.16026}
\showURL{%
\tempurl}


\bibitem[Saeed et~al\mbox{.}(2024)]%
        {saeed2024utilities}
\bibfield{author}{\bibinfo{person}{Ajmal Saeed}, \bibinfo{person}{Mike Jensen}, \bibinfo{person}{Matthew~J Reno}, \bibinfo{person}{Ali Bidram}, {and} \bibinfo{person}{Amin Zamani}.} \bibinfo{year}{2024}\natexlab{}.
\newblock \bibinfo{booktitle}{\emph{Utilities Perspective on Protection Challenges with High IBR Penetration}}.
\newblock \bibinfo{type}{{T}echnical {R}eport}. \bibinfo{institution}{Pacific Gas and Electric}.
\newblock


\bibitem[staff(2021)]%
        {NERC_Odessa_Disturbance}
\bibfield{author}{\bibinfo{person}{NERC staff}.} \bibinfo{year}{2021}\natexlab{}.
\newblock \bibinfo{title}{Odessa Disturbance}.
\newblock \bibinfo{howpublished}{\url{https://www.nerc.com/pa/rrm/ea/Documents/Odessa_Disturbance_Report.pdf}}.
\newblock
\newblock
\shownote{Accessed: 2025-08-10}.


\bibitem[Strohmaier et~al\mbox{.}(2025)]%
        {Top500_2025jun}
\bibfield{author}{\bibinfo{person}{Erich Strohmaier}, \bibinfo{person}{Jack Dongarra}, \bibinfo{person}{Horst Simon}, {and} \bibinfo{person}{Martin Meuer}.} \bibinfo{year}{2025}\natexlab{}.
\newblock \bibinfo{title}{Top500 List June 2025}.
\newblock \bibinfo{howpublished}{\url{https://www.top500.org/lists/top500/2025/06/}}.
\newblock
\newblock
\shownote{Accessed: 2025-08-11}.


\bibitem[{Zachary Skidmore}(2025)]%
        {Skidmore_Apple_renewable_data_center_power}
\bibfield{author}{\bibinfo{person}{{Zachary Skidmore}}.} \bibinfo{year}{2025}\natexlab{}.
\newblock \bibinfo{title}{{Apple data centers consumed more than 2.5 billion kWh over 2024}}.
\newblock
\urldef\tempurl%
\url{https://www.datacenterdynamics.com/en/news/apple-data-centers-consumed-more-than-25-billion-kwh-over-2024/}
\showURL{%
\tempurl}
\newblock
\shownote{Last accessed 2024 August 12}.


\end{thebibliography}

\end{document}